\documentclass[12pt]{article}
\usepackage{amsmath,amsthm,amsfonts,amssymb,amscd}
\usepackage{graphics}
\usepackage[latin1]{inputenc}

\renewcommand{\thefootnote}

\begin{document}

\centerline{\large\bf $\pmb{QCD(SU(\infty))}$ as a model of infinite dimensional}

\centerline{\large\bf  constant gauge field configurations}

\vglue .4in

\centerline{\large\bf Luiz C.L. Botelho}

\vglue .3in

\centerline{Departamento de Matemática}

\smallskip

\centerline{Universidade Federal Fluminense,}

\smallskip

\centerline{Rua Mario Santos Braga, sn, Niterói ,}

\smallskip

\centerline{Rio de Janeiro, zip code 24.220-006, Brazil}

\smallskip

\centerline{e-mail: botelho.luiz@superig.com.br}

\vglue 1in

\noindent
{\bf Abstract:} We study and clarify in a reduced dynamical model for $QCD (SU(\infty))$ called Bollini-Giambiagi model and defined by constant gauge fields Yang-Mills path integral, several concepts on the validity of string representations on $QCD (SU(\infty))$ and the confinement problem.

\vskip .5in

\noindent{\large\bf 1.\, Introduction}

\vglue .2in

In the last decades (since 1980's years), approaches have been pursued to reformulate non supersymetric quantum chromodynamics as a String Theory ([1], [2], [3]) and thus handle the compound hadron structure in the $QCD$ model for strong interactions ([3]). The common idea of all those attempts is to represent the full quantum ordered non supersymetric phase factor as a string path integral, which certainly takes into account more explicitly the geometrical setting of the non abelian gauge theory than its usual description by gauge potential.

\medskip

Other main protocol to achieve such string representation for the wilson loop operator in $QCD$ is to use the still not completely understand large number of colors of t'Hooft for non supersymetric quantum Yang-Mills theory.

\medskip

It is the purpose of this paper to evaluate the static potential between two static charges with opposite signal on the approach of an effective reduced quantum dynamics of Yang-Mills constant-gauge fields ([3]) these results surely are expected to be relevant for the validity of the old conjecture of E. Witten about a $QCD(SU(\infty))$ dynamics dominated by a infinite dimensional ``matrix" constant gauge $SU(\infty)$ master field configuration ([4], [5], [6]).
 These studies are presented in section 2 of this paper. In section 3, we present the relevant $QCD(SU(\infty))$ loop wave equation for our reduced model of constant -- gauge fields for $QCD(SU(\infty))$ and suggest  that, a free bosonic string as solution for this reduced Loop Wave Equation ([7]). We continue with our study and present also a detailed calculation of the quark-antiquark static potential from a one-loop approximation on the Regge slope string constant directly from the well-known Nambu-Goto string path integral ([8]).
\footnote{It is worth to call attention that our constant gauge fields at $SU(\infty)$ are not the rigorous continuum version of the one-plague the Eguchi-Kawai lattice model.}

Finally in section 4, we present also somes studies on the dynamical aspects of this framework of constant field model by presenting path integral studies on evaluation of vectorial-scalar color singlet quark currents ([6],[8]).

Before to proceed, let us firstly reproduce  two enlightneen discourses on the present day problem of handle quantitatively Yang-Millls fields outside the lattice approximation 1-Quoted from A. Jaffe and Eduard Witten.

``Classical properties of non abelian gauge theory are within the reach of established mathematical methods, and indeed, classical non abelian gauge theory has played a very important role in pure mathematics in the last twenty years, especially in the study of three- and four-dimensional ($C^\infty$-differentiable) manifolds.

On the other hand, one does not yet have a mathematically complete example of a quantum gauge theory in four-dimensional space-time, not even a non abelian quantum gauge theory in four-dimensional.

Related to the pure string holographic approach based on the Maldacena conjecture and Super String Theory it appears interesting to cite V. Rivasseu (Math-ph/0006017) about the general philosophy underlying supersymmetric strings.

Today the main strean of theoretical physics holds the view that field theory is only an effective (approximated!) theory and that superstring or its variant, $M$-theory are the best candidate for a fundamental global theory of nature (including $QCD$).

However this superstring theory has not yet received direct experimental confirmation; it has (surely) opened up a new interface with mathematicians, mostly centered around concepts and ideas of geometry and topology (of $C^\infty$-manifolds), with algebra and geometry dominating over analysis and calculational aspects.

Fortunately there is Lattice Gauge Theory, which although has remained largely phenomenological, it has produced somewhat ``precise'' results on Experimentall Hadron mass spectroscopy as it has been pointed out by F. Wilczek (Nature 456,449, 2008). Being enough for that by just taking some mesons $(\pi, \mu, \Sigma)$ mass inputs, even if in the context of $QED$ one needs as input only the fine structure $\alpha = \dfrac{e^2}{4\pi{\hbar} c} \sim \dfrac{1}{137}\cdot$"

In this paper we propose to implement the $QED$ one universal protocol above pointed out by taking now our reduced model as the $QCD$ effective theory at large $N_c$ and as universal imput parameter, the Gluonic condensate $\langle 0|{\rm Tr}(F^2)|0\rangle_{SU(\infty)}$ ([7]), instead of the fine structure constant $\alpha = \dfrac{e^2}{4\pi{\hbar} c}$.

\vglue .3in

\noindent{\large\bf 1.\, The Static Confining Potential for the Bollini-Giambiagi}

\noindent{\large\bf\quad\, model on $D=4$.}

\vglue .3in

The basic gauge-invariant observable on probing the non-perturbative vacuum ([1]) of $SU(N)$ Euclidean Yang-Mills bosonic field theory on $R^4$ is the Wilson loop quantum average
\begin{equation}
\langle W[C]\rangle = \frac{\int d\mu[A]\, W[C]}{\int d\mu[A]} \tag{2.1}
\end{equation}
where the loop parallel transport $SU(N)$-valued matrix is given by
\begin{equation}
W[C] = \frac 1N {\rm Tr}_{SU(N)} \mathbb{P}\left\{\exp \left[ig \oint_C A_\mu(x)\,dx_\mu\right]\right\} \tag{2.2}
\end{equation}
and $d\mu[A]$ denotes the Yang-Mills path-integral measure given formally by the Feynman prescription
\begin{equation}
d\mu[A] = \left(\prod_{x\in R^4} (dA(x))\right) \times \exp\left(-\frac 14 \int_{R^D} {\rm tr}(F_{\mu\nu}^2(A))(x)d^4x\right). \tag{2.3}
\end{equation}

The Gauge connection $A_\mu(x)$ and the field strenght $F_{\mu\nu}(x)$ are explicitly given by
\begin{align*}
&F_{\mu\nu}(x) = \big(\partial_\mu\,A_\nu - \partial_\nu\,A_\mu + ig[A_\mu,A_\nu]_-\big)(x)\\
&A_\mu(x) = A_\mu^a(x)\lambda_a \tag{2.4}
\end{align*}
The $SU(N)$ generators $\big\{\lambda_a\big\}_{a=1,\dots,N^2-1}$, are supposed to be Hemiteans and satisfying the will known matrix relationship below
\begin{align*}
&\big[\lambda^a,\lambda^b\big] = if_{abc}\,\lambda_c\\
&{\rm Tr}\big(\lambda_a\,\lambda_b\big) = \frac{\delta_{ab}}{2}\\
&f^{abc}\,f^{dbc} = N\,\delta^{cd} \tag{2.5}
\end{align*}

In our proposal for the Euchi-Kawai model on continuum, we introduce the space-time trajector $C_{(R,T)}$ of a static quark-antiquark pair, separated apart a distance $R$ with a (Euclidean) temporal evolution $0 \le x_0 \le T$\,\, $(R^4 = \{(x_0,\overset{\rightharpoonup}{x}), \overset{\rightharpoonup}{x} \in R^3)$.

We face thus the problem of evaluating the path integral eq(2-1) for constant gauge fields configurations at the large $N_c$ limit ([5], [6]) and at the physical space-time $R^4$.

Let us briefly review our previous framework on our $D=4$ highly non-trivial generalization of $D=2$ model of Bollini-Giambiagi ([3]).

We firstly consider the full infinite volume space-time $R^4$ reduced to a finite volume space-time $\Omega$. This step has the effect to turns our ``reduced'' path integrals mathematically well defined.

This finite volume space-time is supposed to be formed by the supperposition of $p$ four-dimensional hypercubes of caracteristic volume (``size'') $V = L^4$.

The area $S[C_{(R,T)}]$ enclosed by our rectangle $C_{(R,T)}$ is such that $S[C_{R,T)}] \sim q\,L^2$ for large $p$. Obviously $\big(S[C]\big)^2 \le {\rm vol}(\Omega)$. So ours integers $q$ and $p$ should satisfy $q \le \sqrt{p}$, an important bound to be kept on mind on what follows.

Our improved large $N$-limit will be defined in such way by already taking into account the basic phenomena of $QCD$-Yang-Mills dimensional transmitation of the strong coupling constant, a fully non perturbative phenomena similar to the Higgs mass mechanism on Weinberg-Salan theory. We, thus, define the effective $SU(\infty)$ coupling constant through the relationship for our space-time of finite volume
\begin{equation}
\lim_{N\to\infty} \left(\frac{g^2N}{L^2}\right)\left(\frac qp\right)  < \infty. \tag{2.6}
\end{equation}

It is worth observe that for $p \to \infty$ the infinite volume limit  should be taken according the underlying $N_c\to\infty$ limit. Namely
\begin{equation}
\frac qp \equiv a\to 0 \tag{2.7}
\end{equation}
\begin{equation}
\lim_{\substack{N\to\infty\\ a\to0}} g^2 \left( \frac{Na}{L^2} \right) =(g_\infty)_{\dim}^2 < \infty
 \tag{2.8}
\end{equation}
Here $L^2$ is a physical finite area parameter somewhat to be related to the domain area size of the famous $QCD$ spaghetti vacuum ([7]), a topic to be discussed elsewhere.

After these preliminaries remarks, we must solve the invariant constant gauge field $SU(\infty)$ matrix path integral below written:
\begin{align*}
W_{SU(\infty)}\big[C_{(R,T)}\big] &=\bigg\{\lim_{N\to\infty} \frac{1}{W(0)}\times \bigg[\int_{-\infty}^{+\infty} \left(\prod_{a=1}^{N^2-N} \prod_{\mu=0}^{D-1} d\,A_\mu^a\right)\bigg] \times \\
&\times \Delta_{Fp}[A_\mu]\,\exp\left(+ \frac{g^2}{2}\, V\,{\rm Tr}\bigg(\big[A_\mu, A_\nu\big]_-^2\bigg)\right) \times\\
&\times \left(\exp\, \frac{|g^2S[C_{(R,T)}]|^2}{2} \cdot \frac 1N \, {\rm Tr}_{SU(N)}
\,\bigg([A_0,A_1]^2\bigg)\right)\bigg\}\cdot \tag{2.9}
\end{align*}

In order to evaluate the $SU(N)$-invariant constant Gauge field path integral eq(2.9), we use the Bollini-Giambiagi Cartan matrix decomposition ([5])
\begin{equation}
A_\mu = B_\mu^a\,H_a + G_\mu^b\,E_b \tag{2.10}
\end{equation}
where the Cartan basis $\{H_a,E_a\}$ of the $SU(N)$ Lie algebra possesses the special calculations properties ([5], [6])

\medskip

\noindent a)\, For $a,b = 1,2,\dots,N-1$
\begin{equation}
\big[H_a, H_b\big]_- = 0. \tag{2.11}
\end{equation}

\smallskip

\noindent b)\, For $b = \pm 1,\dots, \pm\,\dfrac{N(N-1)}{2}$
\begin{equation}
\big[H_a, E_b\big]_- = r_a(b)\, E_b. \tag{2.12}
\end{equation}

\smallskip

\noindent c)\, For $a=1,2,\dots,\dfrac{N(N-1)}{2}\cdot$
\begin{equation}
\big[E_a,E_{-a}\big] = \sum_{\ell=1}^{N-1} r_c(a)\, H_a. \tag{2.13}
\end{equation}

\smallskip

\noindent d)\, For $a \ne -b$, $a,b = \pm 1,\dots,\pm\,\dfrac{N(N-1)}{2}$
\begin{equation}
\big[E_a, E_b\big]_- = N_{ab}\,E_{a+b}. \tag{2.14}
\end{equation}

\smallskip

In this distinguished Lie Algebra basis, one can easily fix the Gauge on the $SU(N)$-valued invariant matrix path integral by simply choosing all the $N$-abelian components $B_\mu^a$ on the connection eq(2-10) to be vanished. Namely
\begin{equation}B_\mu^a = 0. \tag{2.15}
\end{equation}

It is  expected thus that the Faddev-Popov term $\Delta_{Fp}[A_\mu]$ should be quenched at the $N \to \infty$ limit eq(2-6)-eq(2-8) i.e. ([5])
\begin{equation}
\lim_{N\to\infty}\,\Delta_{Fp} [A_\mu] \to 1 \tag{2.16}
\end{equation}

Any way we take the Faddev-Popov quenched determinant to unity in this sort of approximate evaluation of ours. $SU(\infty)$-matrix valued invariant path integral (note that procedure to evaluate degrees of Freedom reduced path integrals is usually implemented when one handles for instance, fermions degrees of Freedon on $SU(N)$ lattice path integral ([3]). We will adopt such procedure here).

By assembling all the above results one gets the following outcome eq(2.9) defined now by $SU(N)$ constant gauge field configurations for a general euclidean space-time $R^D$ from now on
$$
W\big[C_{(R,T)}\big] = \frac{1}{W[0]} \times \left\{\int_{-\infty}^{+\infty} \left[\prod_{a=1}^{N^2-N} \prod_{\mu=0}^{D-1} dG_\mu^a\right]\right\}
$$
\begin{equation}
\exp\left\{+ \frac 12\,G_\mu^a\,G_\nu^b\,G_\mu^c\, G_\nu^d\,\mathcal{L}_{abcd} \left[g^2V + \left(\delta_{\mu 0}\,\delta_{\nu 1} \frac{[g^2S][C_{(R,T)}]}{\big(\frac N2\big)}\right)\right]\right\} \tag{2.17}
\end{equation}

\footnote{
\begin{align*}
\mathcal{L}_{abcd} &= \left(\sum_{i,\ell=1}^{N-1} r_i(a)\, r_\ell(c)\,\delta_{i\ell}\,\delta_{c_i-d}\,\delta_{c_i-b}\right)\\
&\quad +\left(N_{ab}\,N_{cd}\,\big(1-\delta_{a_i-b}\big)\big(1-\delta_{c_i-d}\big) \delta_{a+b,-(c+d)}\right)
\end{align*}
}

The above matrix valued path integral can be easily exactly evaluated through re-scalings, at large $N$: Namelly (see Appendix A for details).

\newpage

\noindent a)\, For $\mu \ne 0,1$:\quad $G_\mu^a \to G_\mu^a[g^2V]^{-\frac 14}$\,.

\medskip

\noindent b)\, For $\mu=0,1$;
\begin{equation}
G_\mu^a \to G_\mu^a \left[g^2 +\frac{[g^2S[C_{(R,T)}])^2]}{\big(\frac N2\big)}\right]^{-1/4}\tag{2.18}
\end{equation}
And leading thus to the exactly result
$$
W\big[C_{(R,T)}\big] = \lim_{N\to\infty} \bigg\{\bigg[\bigg(g^2V + \dfrac{(g^2S[C_{(R,T)}])^2}{\dfrac N2}\bigg)^{-\big(\dfrac{N^2-N}{2}\big)}
$$
\begin{equation}
\times\quad \big(g^2V\big)^{-\dfrac{[N^2-N)(D-2)}{4}}\bigg]\bigg/(g^2V)^{-\dfrac{(N_2-N)D}{4}}\bigg\} \tag{2.19}
\end{equation}

It yields thus, the following $N \to \infty$ limit
$$
W\big[C_{R,T}\big] = \lim_{N\to\infty} \left(1+\frac{g^2S^2\big[C_{R,T)}\big]}{\bigg(\dfrac N2\bigg)V}\right)^{\dfrac{-N(N-1)}{2}}
$$
\begin{equation}
=\quad \lim_{N\to\infty} \left\{\exp\left[-\frac{g^2(N-1)L^2}{L^2}\,\left(\frac{S^2}{V}\right)\right]\right\} \tag{2.20}
\end{equation}

For $D=4$, we have on the context of our proposed $SU(\infty)$ infinite volume limit eqs(2-6)-(2-8), our $R^4$ Wilson Loop ``string'' behavior.
\begin{align*}
W\big[C_{(R,T)}\big] = \exp\left\{- \frac{g^2(N-1)}{L^2} L^2 \left(\frac{q^2L^4}{pL^4}\right)\right\}
& \overset{(N\to\infty)}{\sim}\quad   \exp\left\{-\frac{g^2N}{L^2}\,\left(\frac qp\right) (qL^2)\right\}\\
 & \overset{(N\to\infty)}{\sim}\quad \exp\left\{-(g_\infty)^2 \,RT\right\} \tag{2.21}
\end{align*}

\noindent From eq(2-21), one obtains the confining quark-antiquark potential
\begin{equation}
V(R) = \lim_{T\to\infty}\,\, \left(- \frac 1T\, \ell g \big(W\big[C_{R,T)}\big]\big)\right) = (g_\infty)^2\,R \tag{2.22}
\end{equation}
leading to an attractive constant force ``biding'' the static pair of quarks as originally obtained by K. Wilson on his lattice gauge - modelling ([3]).

\newpage

\noindent{\large\bf 3.\, The Luscher correction to inter quark potential on the}

\noindent\qquad{\large\bf reduced model}

\vglue .3in

In this section we intend to show that our proposed $SU(\infty)$ constant gauge field theory leads to a free string theory path-integral. We thus  evaluate explicitly through the string path-integral the next non-confining corrections to the quark-antiquark potential eq(2-22).

\medskip

In order to argument an effective low energy $QCD$ string representation in this model, we are going to consider the loop have equation ([1]) for constant gauge fields already on the continuum at large $N$ limit.

\medskip

Let us thus firstly consider general loops $C_{x\,x} = \big\{\big(X_\mu(\sigma)\big)_{\mu=0,1,2,3}\,; \, 0 \le \sigma \le 2\pi\big\}$ on $R^4$. It is well-known that formally we have the functional loop derivative ([1])

\smallskip

\noindent a)

\begin{equation}
\psi\big[C_{x,x},A_\mu(x)\big] = \frac 1N\, \text{Tr } \mathbb{P}\,\left\{\exp \big(ig \oint_{C_{xx}}\,A_\mu(X(\sigma))dX^\mu(\sigma)\big)\right\} \tag{3.23}
\end{equation}

\noindent b)

\begin{align*}
\frac{\delta}{\delta X_\mu(\overline \sigma)} \quad \psi\big[C_{xx},A_\mu(x)\big] &= 
\frac{ig}{N}\, \text{Tr } \mathbb{P}\bigg\{F_{\mu\nu} \big(X_x(\overline{\sigma})\big) \frac{dX^\nu(\overline\sigma)}{d\overline \sigma}\bigg)\\
&\quad \times \exp\left(ig \oint_{C_{xx}} A_\mu(X(\sigma))dX^\mu(\sigma)\right)\bigg\}, \tag{3.24}
\end{align*}

\noindent c)
$$
\frac{\delta^2}{\delta X_\mu(\overline\sigma)\delta X^\mu(\overline\sigma)}\quad \psi_{SU(N)} \, \big[C_{xx}, A_\mu(x)\big] =
$$
$$
=\bigg\{\bigg[\frac 1N\, ig \text{ Tr } \mathbb{P} \bigg\{\overbrace{(\nabla F_{\mu\nu})(X^\beta(\overline\sigma)) \frac{dX^\nu}{d\overline \sigma}(\overline \sigma)}^{=\dfrac{\delta}{\delta X_\mu(\overline\sigma)} (F_{\mu\nu}(X(\overline\sigma)))}\,\exp\left(ig \oint_{C_{xx}} A_\mu dX_\mu)\right)\bigg]
$$
\begin{equation}
+\bigg[\frac{(ig)^2}{N}\,\text{ Tr } \mathbb{P} \bigg\{(F_{\mu\nu} F_{\mu\nu})(X^\beta(\overline\sigma)) \times 
\left(\frac{dX^\rho}{d\overline\sigma}\cdot \frac{dX^\rho}{d\overline\sigma}\right)(\overline\sigma)\,
\exp\left(ig \oint_{C_{xx}} A_\mu dX_\mu)\right)\bigg]\bigg\}\tag{3.25}
\end{equation}

\medskip

For constant gauge fields configurations the first term of the right-hand side of eq(3.25) $\bigg(\dfrac{\delta}{\delta X_\mu(\overline\sigma)} (\text{ constant } F_{\mu\nu})=0!\bigg)$ vanishes identically. So, after taking the path integral average of eq(3.25) through the path integral of constant gauge fields configurations eq(2.9) and considering the usual path integral factorization of a product of gauge invariant observable at $SU(\infty)$, together with the formation of non-vanishing value of the Yang-Mills energy on the non-trivial $QCD$ vacuum one gets finally the following loop wave equation for the quantum Wilson Loop  our Loop in our reduced $SU(\infty)$ gauge theory on $R^4$. 
$$
\int_0^{2\pi}\left[\left(\frac{\delta^2}{\delta X_\mu(\overline\sigma)\,\delta X^\mu(\overline\sigma)}\right)\, \psi_{SU(\infty)} \big[X^\mu(\sigma)]\right] d\overline\sigma
$$
$$
(\ge 0)
$$
\begin{equation}
= \left(-\big(g_\infty\big)_{(0)}^2\, \langle 0|F^2|0\rangle_{SU(\infty)}\right) \times \left(\int_0^{2\pi}\big\vert X_\mu^{\prime}(\overline\sigma)\big\vert^2 \times \psi_{SU(\infty)} \big[X^\mu(\sigma)]\big] d\overline\sigma\right) \tag{3.26}
\end{equation}
 
\smallskip

\noindent Here, we have the $SU(\infty)$ Euclidean Gauge Theory parameters identification 

\smallskip

\noindent a)$^*$\footnote{*-\,\, Gerard t'Hooft hypothesis}
$$
\big(g_\infty\big)_{(0)}^2= \lim\limits_{N\to\infty}\,(g^2N) < \infty
$$ 

\smallskip

\noindent b)$^{**}$\footnote{**- One of the main points on the search for string representations on Q.C.D. $(SU(\infty))$ (formally a quantum field theory defined by only by ``planar" Q.C.D., infrared regularized, Feynman diagramnas - G. t'Hooft), is the the non-perturbative condensate formation of the Yang-Mills $SU(\infty)$ field strenght:
$$
\lim_{x\to x'} \big\langle \Omega_{vac}^\infty\,\big\vert \big(F_{\mu\nu}(x)\,F^{\rho\sigma}(x')\big)\big\vert \Omega_{vac}\big\rangle = \delta^{\mu\rho} \delta^{\nu\sigma} \big\langle \Omega_{vac}^\infty \big\vert F^2(x)\big\vert \Omega_{vac}^\infty \big\vert.
$$

This hypothesis has the same foundational importance of the non-zero formation of the Higgs field expectation value on Weinberg-Salan Weak interaction theory.

The (non-perturbative) non vanishing of eq(3.27) is a fundamental hypothesis on all of ours works on the subject ([3]).}

\begin{equation}
\delta^{\rho\alpha}\,\langle 0|F^2|0\rangle_{SU(\infty)} = \lim_{N\to\infty} \left(\frac 1N\,\text{Tr } \big\langle 0\big\vert \int d^D\,x\big(F_{\mu\rho}(x)\,F^{\mu\alpha}(x)\big)\big\vert 0\big\rangle\right) < 0 \tag{3.27}
\end{equation}

\smallskip

\noindent By comparing the above parameters with those coupling constants of the static case,one has the following identification for the Spaghetti $QCD$ non perturbative broken scale invariance vacuum effective area domain with the $QCD$ value condensate
\begin{equation}
\frac{1}{a^{\rm eff}} = \big(-\big\langle 0|F^2|0\big\rangle\big) > 0. \tag{3.28}
\end{equation}
A result already expected ([7]).

\medskip

At this point we point out that the reduced loop wave equation is the same of a free Bosonic string theory with the string Regge slope identification with the reduced Gauge theory at $SU(\infty)$
\begin{equation}
\frac{1}{(2\pi\alpha')^2} = -\big(g_\infty\big)^2_{(0)} \langle 0 |F^2| 0\rangle \tag{3.29}
\end{equation}

As a consequence, one should expects the phenomenological path-integral representation between the large $N_c$ and extreme low energy continuum $QCD(SU(\infty))$ (represented by constant $SU(\infty)$ gauge fields), with a free bosonic (creation process) string path integral on the light-cone gauge$^{(*)}$\footnote{(*)- Rigorously, one should expect the Polyakov's covariant path integral for $D=4$ (with Liouville quantum degrees of freedon as representing the QCD $(SU(\infty))$ Wilson Loop ([8]).}
$$
\big\langle\psi\big[C_{xx},A_\mu(x)\big]\big\rangle_{\substack{SU(\infty)\\ \rm low-energy}} =G(C_{xx},A)
$$
$$
= \int_0^\infty dA\bigg\{\int_{X^\mu(\sigma,0)=0}^{X^\mu(\sigma,A)=C_{xx}(\sigma)} D^F[X^\mu(\sigma(\zeta)]\bigg\} 
$$
\begin{equation}
\,\times \exp\bigg\{-\frac 12 \int_0^A dt \int_0^{2\pi} d\sigma\left[\big(\partial_\zeta X^\mu\big)^2 + \frac{1}{(2\pi\alpha')^2} \big(\partial_\sigma X^\mu\big)^2\right]\bigg\} \tag{3.30}
\end{equation}

\smallskip

\begin{equation}
G_{\rm string} \big(C_{xx},0\big) = G_{\rm string} \big(C_{xx},\infty\big) = 0 \tag{3.31}
\end{equation}

\smallskip

At this point it is worth observe that our light-cone string path integral propagator
$$
G_{\rm string} \big(C_{xx},A\big) = \int_{X^\mu(\sigma,0) = C_{xx}(\sigma)}\, D^F[X^\mu(\sigma,\pi)]
$$
\begin{equation}
\times\, \exp\bigg\{-\frac 12 \int_0^A dt \int_0^{2\pi} d\sigma \left[\big(\partial_\zeta X^\mu\big)^2 + \frac{1}{(2\pi\alpha')^2}\,\big(\partial_\sigma\,X^\mu\big)^2\right]\bigg\} \tag{3.32}
\end{equation}

\smallskip

\noindent satisfies the area-Difusion euclidean Schrodinger loop functional equation:
\begin{equation*}
\frac{\partial G_{\rm string}\big(C_{xx},A\big)}{\partial A} = \left\{\int_0^{2\pi} d\overline{\sigma} 
\left(\frac{\delta^2}{\delta X_\mu(\overline\sigma)\delta X_\mu(\overline\sigma)} - \frac{1}{(2\pi\alpha')^2} |X_\mu'(\overline\sigma)|\right) G_{\rm string} \big(C_{xx},A\big) \right\}\hfill(3.33)
\end{equation*}

\smallskip

\noindent togheter with the boundary conditons:
\begin{equation}
G_{\rm string} \big(C_{xx},0\big) = G_{\rm string} \big(C_{xx},\infty\big) = 0. \tag{3.34}
\end{equation}

\medskip

Let us now evaluate in details the quark-antiquark potential from the general Nambu-Goto string, path integral in $R^D$
\begin{equation}
\psi\big[C_{(R,T)}\big] = \int D^F\big[X^\mu(\sigma,\zeta)\big] \times \exp\left\{-\frac{1}{2\pi\alpha'} \left[\int_0^T d\zeta \int_0^R d\sigma\,\sqrt{\det(h(X^\mu(\sigma,\zeta))}\right]\right\} \tag{3.35}
\end{equation}

\medskip

Here the orthogonal dynamical string vector position is considered as closed quantum fluctuations from the static quark-antiquark trajected by $C_{(R,T)}$ i.e.:

\smallskip

\noindent a)
$$
X^\mu(\sigma,\zeta) = \zeta(1,0) + \sigma(0,1) + \sqrt{\pi\alpha'}\,Y^\mu(\sigma,\zeta)
$$

\smallskip

\noindent b)
$$
\begin{cases}
Y^\mu(\sigma,\zeta\pm T) = Y^\mu(\sigma,\zeta)&\\
Y^\mu(0,\pm T) = Y^\mu(0,0)&\\
\mu = 2,3,\dots,D-2&
\end{cases}
$$

\smallskip

\noindent c)
$$
h_{00}\big(X^\mu(\sigma,\zeta\big) = \big(\partial_\zeta X^\mu\, \partial_\zeta X_\mu\big)(\sigma,\zeta) = 1 + \pi\alpha'\big(\partial_\zeta Y^\mu\, \partial_\zeta Y^\mu\big) (\sigma,\zeta)
$$

\smallskip

\noindent d)
$$
h_{01}\big(X^\mu(\sigma,\zeta\big) = \pi\alpha'\big(\partial_\sigma Y^\mu\, \partial_\zeta X^\mu\big)(\sigma,\zeta)
$$

\smallskip

\noindent e)
\begin{equation}
h_{11}\big(X^\mu(\sigma,\zeta\big) = 1 + \pi\alpha'\big(\partial_\sigma Y^\mu\, \partial_\sigma X^\mu\big) (\sigma,\zeta) \tag{3.36}
\end{equation}

\smallskip

As a consequence, we have explicitly the following one-loop order approximation for the string path integral weight eq(3.35):
\begin{equation}
\frac{1}{2\pi\alpha'} \sqrt{h\big(X^\mu(\sigma,\zeta)}\big) = \frac{1}{2\pi\alpha'} \left[1+\pi\alpha'\left(\frac{\partial Y^\mu}{\partial\zeta}\, \frac{\partial Y^\mu}{\partial\zeta} + \frac{\partial Y^\mu}{\partial\sigma}\,\frac{\partial Y^\mu}{\partial\sigma}\right)\right] \times 0\big((\alpha')^2\big) \tag{3.37}
\end{equation}

\medskip

As a result of substituting eq(3.36) on eq(3.37), one gets the following closed bosonic string Gaussian path-integral to evaluate
\begin{align*}
&\psi\big[C_{(R,T)}\big] =\\
&\int_{Y^\mu(\sigma,\zeta\pm T)=Y^\mu(\sigma,\zeta)}\, D^F\big[Y^\mu(\sigma\zeta)\big] \, \exp\left\{\left(-\frac{RT}{2\pi\alpha'}\right) - \frac 12 \int_0^T d\zeta \int_0^R d\sigma\big(Y^\mu\big(-\Delta\big)_{(R,T)}'\big)Y_\mu\big)(\sigma,\zeta)\right\}\\
&= e^{-\frac{RT}{2\pi\alpha'}} \left(\det{}^{-\frac{(D-2)}{2}} \big(-\Delta\big)_{(R,T)}^\prime\bigg)\right) \tag{3.38}
\end{align*}

\smallskip
\noindent where the Laplacian $-\Delta_{(R,T)}^\prime$ on the rectangle $C_{(R,\pm 1)}$ has Dirichlet boundary conditions and considered projected out from the zero modes.

\medskip

It has been evaluated fully on the literature ([8]):
\begin{align*}
&\det{}^{-\frac{D-2}{2}} \big(-\Delta_{(R,T)}^\prime\big) 
= \left[\left(\frac RT\right)^{(\frac{D-2}{2}}\,e^{+\frac{\pi T}{6R}(D-2)} \times 
\left(\prod_{n=1}^\infty \left(1-e^{-\frac{2\pi n}{R}(\frac TR)}\right)^{-2(D-2)}\right)\right] \tag{3.39}
\end{align*}

\medskip

At this point, one can easily verify the string result for the quark-antiquark potential with the Coulomb interaction Lüscher correction.
\begin{equation}
V(R) = \lim_{T\to\infty} \left(-\frac 1T\, \ell n\,\psi\big[C_{(R,T)}\big]\right) = \frac{1}{2\pi\alpha'}\,R - \frac{(D-2)\pi}{6}\cdot \frac 1R \tag{3.40}
\end{equation}

\vglue .3in

\noindent{\large\bf 4.\, Some path integral dynamical aspects of the reduced $QCD$}

\noindent\qquad{\large\bf as a path integral dynamics of euclidean strings}

\vglue .3in

Let us start this section on dynamical aspects by writing firstly in details, the operational euclidean path integral expression for the non-relativistic Feynman propagator of a spinless particle in the presence of an external (euclidean) abelian gauge field and an external scalar potential.

\medskip

As a first step let us write the Feynman propagator above cited in the euclidean space-time $R^4(\hbar = 1)$
\begin{equation}
G(x,y,t) = \big\langle x \left\vert \exp\left\{-t \left[\frac{1}{2m}\,\big(-i\overline{\nabla} - 
\frac ec\,\vec{A}\big)^2 - i\,e\,\varphi + g\,V\right]\right\}\right\vert y\big\rangle \tag{4.41}
\end{equation}

\medskip

Then $\vec{A} = (A_1,A_2,A_3)$ denotes the time-dependent vectorial abelian field, $\varphi$ the field potential and $V(x,,t)$, the external potential, also supposed time dependent.

\medskip

The phase-space path integral is easily written as of as
\begin{align*}
&G(x,y,t) = \int_{\vec{X}(0)=y}^{\vec{X}(t)-x} D^F(\overset{\rightarrow}{X}(\sigma)) \int D^F[\overset{\rightarrow}{p}(\sigma)]\\
&\times\,\exp\left\{+i \int_0^t \left(\overset{\rightarrow}{P}\cdot\frac{d\overset{\rightarrow}{A}}{d\sigma}\right)(\sigma)\,d\sigma\right\}\\
&\times \exp\left\{-\left[\int_0^t \frac{1}{2m}\,\left(\overset{\rightarrow}{P}(\sigma) - \frac{e}{c}\,\overset{\rightarrow}{A}(\overset{\rightarrow}{X}(\sigma))\right)^2 - i\,e\,\varphi(\overset{\rightarrow}{X}(\sigma),\sigma) + g\,V(X(\sigma),\sigma)\right]\right\} \tag{4.42}
\end{align*}

\medskip

After the formal path-integral change of variable into eq(4.42)
\begin{equation}
\overset{\rightarrow}{P}(\sigma) - \frac{e}{e}\,\overset{\rightarrow}{A}(\overset{\rightarrow}{X}(\sigma),\sigma) = \overset{\rightarrow}{Q}(\sigma) \tag{4.43}
\end{equation}
we get the following result:
\begin{align*}
G(x,y,t) &= \int_{\overset{\rightarrow}{X}(0)=y}^{\overset{\rightarrow}{X}(t)=x} D^F(\overset{\rightarrow}{X}(\sigma)\, \int D^F[\overset{\rightarrow}{Q}(\sigma)]\\
&\exp\left\{i \int_0^t \left(\overset{\rightarrow}{Q}(\sigma) + \frac ec\, \overset{\rightarrow}{A}(X(\sigma),\sigma'\right)\, \frac{d\overset{\rightarrow}{X}(\sigma)}{d\sigma}\right\}\\
&\exp \left\{- \frac{1}{2m}\, \int_0^t (\overset{\rightarrow}{Q}(\sigma))^2 \right\}\\
&\exp \left\{ +ie \int_0^t \varphi(\overset{\rightarrow}{X}(\sigma),\sigma)d\sigma \right\}\\
&\exp \left\{ -\int_0^t V(\overset{\rightarrow}{X}(\sigma),\sigma)d\sigma \right\} \tag{4.44}
\end{align*}

\smallskip

After realizing the Gaussian $\overset{\rightarrow}{Q}(\sigma)$ functional integral
\begin{align*}
\int D^F[\overset{\rightarrow}{Q}(\sigma)]\, &\exp\left[-\frac{1}{2m} \int_0^t \big(\overset{\rightarrow}{Q}(\sigma)\big)^2\,d\sigma\right]\, \exp\left[i \int_0^t \overset{\rightarrow}{Q}(X(\sigma))\,\frac{d\overset{\rightarrow}{X}(\sigma)}{d\sigma}\right] =\\
&\exp\,\left\{-\frac{m}{2}\, \int_0^t \left(\frac{d\overset{\rightarrow}{X}}{d\sigma}\right)^2(\sigma)\,d\sigma \right\}, \tag{4.45}
\end{align*}

\smallskip

\noindent we get our gauge invariant path integral expression for the euclidean Feynman propagator under study 
\begin{align*}
G(x,y,t) &= \int_{\overset{\rightarrow}{X}(0)=y}^{\overset{\rightarrow}{X}(t)=x} D^F[\overset{\rightarrow}{X}(\sigma)]\, \exp\left[-\frac m2 \int_0^t \left(\frac{d\overset{\rightarrow}{X}}{d\sigma}\right)^2(\sigma)\,d\sigma\right]\\
&\exp\left[-g\int_0^t V(X(\sigma),\sigma)\,d\sigma \right]\\
&\exp \bigg[ie\left(\int_0^t \overset{\rightarrow}{A}(\overset{\rightarrow}{X}(\sigma)\cdot \frac{d\overset{\rightarrow}{X}}{d\sigma}\,d\sigma \right)\\
&\qquad + ie \left(\int_0^t \varphi(\overset{\rightarrow}{X}(\sigma),\sigma)\,d\sigma\right) \bigg] 
\end{align*}

\smallskip

It is curious to point out the gauge invariance of eq(4.46) is solely under all periodic Gauge transformation $(x\le z \le y\,\, ; \,\, 0 \le t' \le t)$
\begin{equation}
\begin{cases}
\vec A(z,t') &= (\vec A+ \vec\nabla \Lambda)(z,t') \\
\varphi(z,t') &= (\varphi+\frac{\partial\Lambda}{\partial t}(z,t') \\
\Lambda(x,t) &= \Lambda(y,0)+ \frac{2\pi n}e
\end{cases} \tag{4.47}
\end{equation}

\smallskip

\noindent Namely
\begin{align*}
&\exp \left\{ie \int_0^t (\overline{\nabla}\Lambda)(\overline{X}(\sigma),\sigma)\, \frac{d\overline X}{d\sigma}\,d\sigma + ie \int_0^t \frac{\partial\Lambda}{d\sigma}\,d\sigma\right\}\\
= &\exp \left\{ ie \int_0^t \frac{d}{d\sigma} \big(\Lambda(\overline{X}(\sigma),\sigma)\big) d\sigma\right\}\\
= &\exp \left\{ie\big(\Lambda(x,t) - \Lambda(y,0)\right\}. \tag{4.48}
\end{align*}

\smallskip

In the euclidean quantum field case in $R^D$ one must consider the generating fermionic case
\begin{equation}
Z[A_\mu] = \frac{\det\,^{\frac 12} \big(\not\!D^*(A)\,\not\!D(A)\big)}{\det\,^{\frac 12} (\not\!\partial^*\not\!\partial)}, \tag{4.49}
\end{equation}

\smallskip

\noindent which can be re-write through well-known propertion loop space techniques as a loop space $D$-dimensionall non-relativistic propagator
\begin{align*}
&\ell g\big(Z[A]/Z(A=0)\big) =- \frac 12 \int_0^\infty \frac{dt}{t} \bigg\{\int_{R^D} d^Dx_\mu \int_{X_\mu(0)=x_\mu}^{X_\mu(t)=x_\mu} D^F[X_\mu(\sigma)]\\
&\exp \left\{- \frac{1}{2} \int_0^t \left(\frac{dX^\mu}{d\sigma}\right)^2(\sigma)\,d\sigma\right\}
\\
&\mathbb{P}_{\text{spin}}\bigg\{\mathbb{P}_{SU(N)} \bigg[ \exp \Big( ie \int_0^t\bigg( A_\mu(X^\mu(\sigma))\bigg)
 + \frac{ie}{4}\, [\gamma^\mu,\gamma^\nu]\, F_{\mu\nu}(X^\rho(\sigma)) \,d\sigma\bigg)\bigg]\bigg\}, \tag{4.50}
\end{align*}

\smallskip

\noindent where the symbols $\mathbb{P}_{\text{spin}}$ and $\mathbb{P}_{SU(N)}$ means $\sigma$-ordered matrixes indexes of the spin-color gauge connection phase factor (the fermionic Wilson loop). At very low energy region, one could consider as an effective theory the situation that all degrees of Dirac spin of the particle, are non-dynamical (i.e.: frozen to scalar values), or equivalently one can disregard the spin orbit shenght field coupling on eq(4.50) $\bigg(\dfrac{ig}{\hbar}\,[\gamma^\mu,\gamma^\nu] F_{\mu\nu}(X^\beta(\sigma)) \cong 0\bigg)$.

Let us now apply the above well-known remarks to evaluate approximately ``scalar'' composite operators quark-antiquark Green functions.

The effective connected generating functional for vectorial quark currents at very low energy (the strong coupling region of the underlying Massless Yang-Mills theory) is given by the following loop expression
\begin{equation}
\ell g\left(Z_{QCD}^{eff}\,[J_\mu]\right) = \ell g \bigg\langle \det\,^{\frac 12}\big(\not\!D^*(ig A_\mu + J_\mu) \not\!D(ig A_\mu + J_\mu)\big)\bigg\rangle_{A_\mu} \tag{4.51}
\end{equation}

\smallskip

\noindent there $\langle\,\,\rangle_{A_\mu}$ denotes the complete Yang-Mills path integral, $A_\mu$ the Yang-Mills field and $J_\mu(x)$ the external source of the vectorial quark currents $\big(J_\mu(x)(\overline{\psi}\gamma^\mu\psi)(x)\big)$.

\medskip

On the basis of the above discusions one has the following expression for eq(4.50), with the $QCD$ scale $\Lambda_{QCD}$ already bult in a large $SU(\infty)$ limit in the proper-time gauge (or in the string light-cone gauge)
\begin{align*}
&\ell g\left(Z_{QCD}^{eff} [J_\mu]_{\Lambda_{QCD}}\right) = -\frac 12 \bigg\{\int_{1/\Lambda_{QCD}}^{\Lambda_{QCD}} \frac{dt}{t} \times\\
&\times \bigg[\int d^Dz_\mu \int_{X^\alpha(0)=X^\alpha(t)=z^\alpha} D^F(X^\alpha(\sigma))\, \exp\bigg(-\frac 12 \int_0^t \left(\frac{dX}{d\sigma}\right)^2(\sigma)\,d\sigma\bigg) \times\\
&\times \bigg\langle \mathbb{P}_{SU(N)}\bigg\{\exp\, ig \int_0^t \left(A_\mu\,\frac{dX^\mu}{d\sigma}\right)(\sigma)d\sigma\bigg\}\bigg\rangle_{A_\mu}\,\exp\left(i\int_0^t \left(J_\mu\,\frac{dX^\mu}{d\sigma}\right)d\sigma\right) \tag{4.52}
\end{align*}

\smallskip

The vectorial $N$-point bilinear quark current is given by in momentum space
\begin{align*}
&\big\langle (\overline{\psi} \gamma^{\mu_1}\,\psi)(x_1)\dots(\overline{\psi}\gamma^{\mu_N}\,\psi)(x_\mu)\big\rangle_{A_\mu}\\
&= \frac{\delta^2}{\delta J_{\mu_1}(x_1),\dots,\delta J_{\mu_N}(x_1)} \left[\ell g\left(Z_{QCD}^{eff}(J_\mu)\right)\right] = G_{\mu_1\dots\mu_N}(x_1,\dots,x_N) \tag{4.53}
\end{align*}

Or equivalently, after suitable Fourier momenta transforms.
\begin{align*}
\widetilde{G}_{\Lambda_{QCD}}\big(P_{\mu_1},\dots,P_{\mu_N}\big) &= -\frac 12\,(i)^N \bigg\{\int_{1/\Lambda_{QCD}}^{\Lambda_{QCD}} \frac{dt}{t} \int_0^t d\sigma_1\dots \int_0^t d\sigma_N
\int d^Dz_\mu \\
& \times \int_{X_\mu(0)=z_\mu}^{X_\mu(t)=z_\mu} D^F[X(\sigma)]\\
&\times \bigg[ \exp\left(-\frac 12 \int_0^t \left(\frac{dX}{d\sigma}\right)^2(\sigma)d\sigma\right)\\
&\times \left(\frac{dX_{\mu_1}}{d\sigma}(\sigma_1)\dots\frac{dX_{\mu_N}}{d\sigma} (\sigma_N)\right)\bigg]\\
&\times \left(\exp\left(i \sum_{h=1}^N p_k^\mu\,X_\mu(\sigma_k)\right)\right)\\
&\times \bigg\langle e^{ig \displaystyle\int_0^t A_\mu\,dX^\mu}\bigg\rangle_{SU(\infty)} \tag{4.54}
\end{align*}

\medskip

On the basis of eq(4.54), one could envisage to try to evaluate eq(4.53) through an Gaussian (euclidean) string path integral. Let us take for granted such string representation as a workable sound hypothesis on basis of our previous studies.

\medskip

The key point is to evaluate in terms of the loop variable $X^\mu(\sigma)$, the following anihillation string path integral:\footnote{If the action is $\int_0^A ds \int_0^t d\sigma[(\partial_sY^\mu)^2+ \frac1{(\pi\alpha')^2}(\partial_\sigma Y^\mu)^2](s,\sigma)$, then eq.(4.56) takes the form $\sigma\to \overline\sigma=\sigma(\pi\alpha')$ and $\overline Y^\mu(\overline\sigma,\overline s)=Y^\mu(\sigma,s)$.}
\begin{align*}
&W\big[X_\mu(\sigma),0\le\sigma\le t\big] = \int_0^\infty dA\bigg\{ \int_{Y^\mu(\sigma,0)=X^\mu(\sigma)}^{Y^\mu(\sigma,A)=0} D^F(Y^\mu(\sigma,s))\\
&\qquad \exp \left\{-\frac 12 \int_0^A ds \int_0^t d\sigma \left[\big(\partial_sY^\mu\big)^2 + \frac{1}{(\pi\alpha')}\,\big(\partial_\sigma Y^\mu\big)^2\right](\sigma,s)\right\}\bigg\} \tag{4.55}
\end{align*}

\smallskip

In order to evaluate eq(4.55) exactly, let us firstly consider the stndard re-scale
\begin{align*}
&\sigma \longrightarrow \sigma\big(\pi\alpha'\big)^{1/2} = \overline\sigma\\
&s \longrightarrow s\\
&Y^\mu(\sigma,s) \longrightarrow \overline{Y}^\mu(\overline\sigma,s) \equiv \big(\pi\alpha'\big)^{1/4} (Y^\mu(\sigma,s)) \tag{4.56}
\end{align*}

\smallskip

\noindent which formally turns the string velocity into a overall factor into the path integral weight
\begin{align*}
&W\big[\,\overline{X}_\mu(\overline\sigma)\big] = \int_0^\infty dA \bigg\{\int_{\overline{Y}^\mu(\overline\sigma,0)=\overline X_\mu(\overline\sigma)}^{\overline{Y}^\mu (\overline\sigma,A)=0} 
D^F\big[\overline{Y}^\mu(\overline\sigma,s)\big]\\
&\times \exp\left\{-\frac{1}{(2\pi\alpha')} \left[\int_0^A ds \int_0^{t(\pi\alpha')^{1/2}} d\overline\sigma\left(\left(\frac{\partial\overline{Y}^\mu}{ds}\right)^2 + \left(\frac{\partial\overline{Y}^\mu}{d\overline\sigma}\right)^2\right)\right]\right\} \tag{4.57}
\end{align*}

\smallskip

After considering the ``Brownian Bridge like'' background loop-surface decomposition which has the meaning of considering a toroidal like fluctuating closed string world sheet $Z_\mu(\overline\sigma,\zeta)$ bounded by the closed quark-antiquark trajectory $\overline{X}_\mu(\overline\sigma)$ \big($\overline{X}_\mu(\overline\sigma+t)
= \overline{X}_\mu(\overline\sigma)$, $t$, fixed loop proper-time\big)
\begin{align*}
\overline{Y}_\mu(\overline\sigma,s) &= \overline{X}_\mu(\overline\sigma) \left(\frac{A-s}{A}\right) + \sqrt{\pi\alpha'}\,Z^\mu(\overline\sigma,s)\\
Z^\mu(\overline\sigma,A) &= Z^\mu(\overline\sigma,0) = 0\\
Z^\mu(\overline\sigma+t,s) &= Z^\mu(\overline\sigma,s), \tag{4.58}
\end{align*}

\smallskip

\noindent one gets the regularized proper-time string propagator
\begin{align*}
&W\big[\,\overline{X}_\mu(\overline\sigma)\big] = \int_{\varepsilon}^\infty dA\,\exp\bigg\{-\left(\frac{(A/3)}{2\pi\alpha'}\right)\left(\int_0^{(\pi\alpha')^{\frac 12}\cdot t} \left(\frac{d\overline{X}_\mu(\overline\sigma)}{d\overline\sigma}\right)^2\,d\overline\sigma\right)\\
&- \left(\frac{1}{2\pi\alpha'A}\right)\left(\int_0^{(\pi\alpha')^{\frac 12}\,t} (\overline{X}_\mu(\overline\sigma))^2\,d\overline\sigma\right)\bigg\} \times \left(\det\,^{-\frac D2}\left(-\Delta_{(\pi\alpha')^{\frac 12}\,t,A)}\right)\right) \tag{4.59}
\end{align*}

\smallskip

Just for completeness, we not the following exactly expressions for the fluctucting worl-sheet $Z^\mu$ Laplacean determinant and its Green function on the rectangle\linebreak  $\overbrace{\left[0,(\pi\alpha')^{\frac 12}\,t\right]}^{\sigma} \times \overbrace{[0,A]}^{\xi}$:
\begin{align*}
&\det\,^{-\frac D2}\left(-\Delta_{((\pi\alpha)^{\frac 12} t,A)}\right) = \left(\prod_{n,m} \left[\int \left(\frac{2\pi n}{((\pi\alpha')^{\frac 12}t}\right)^2 + \left(\frac{2\pi m}{A}\right)^2\right]\right)^{-D/2}\\
&= \left(\frac{(\pi\alpha')^{\frac 12}t}{A}\right)^{D/2}\,\,\exp\left(\frac{\pi}{6}  \left(\frac{2\pi nA}{(\pi\alpha')^{\frac 12}\,t}\right) D\right)\\
&\times \left(\prod_{n=1}^\infty \left[1-\exp\left(\frac{2\pi nA}{(\pi\alpha')^{\frac 12}\,t}\right)\right]\right)^{-2D}; \tag{4.60-a}
\end{align*}
$$
\qquad\quad\left(-\Delta\right)_{((\pi\alpha')^{\frac 12}\,t,A)}^{-1}\,\, (\overline\sigma,\overline{\sigma}',s,s') =
$$
$$
\quad = -\frac 12 \bigg\{\sum_{\substack{n,m\\-\infty}}^{+\infty}
\bigg(\frac{e^{\frac{2\pi in(\overline\sigma-\overline\sigma')}{(\pi\alpha')^{\frac 12}\,t}}}
{\left(\frac{2\pi n}{(\pi\alpha')^{\frac 12}\,t}\right)^2 + \left(\frac{2\pi m}{A}\right)^2}\bigg)\times
$$
\begin{equation}
\times \left[\cos\left(\frac{2\pi m}{A} (s-s')\right) - \cos\left(\frac{2\pi m}{A} (s+s')\right)\right]\bigg\}
\tag{4.60-b}
\end{equation}

\smallskip

As a consequence we get for $N$-point euclidean scalar meson Green function after disregarding the contribution of the functional determinant eq(4.60-a) and by considering $\pi\alpha'=1$ from now on
\begin{align*}
&\qquad\quad \widetilde{G}_{(t)}(P_1^\mu,\dots,P_N^\mu)\\
&=-\frac 12 \times 
\bigg\{\int_{1/\Lambda_{QCD}}^{\Lambda_{QCD}} \frac{dt}{t} \int_0^t d\sigma_1\dots \int_0^t d\sigma_N \bigg[ \int_{\varepsilon}^\infty dA \times\\
&\quad \times F\big((P_k^\mu\cdot P_{k'}^\mu),A,T,\{\sigma_1,\dots,\sigma_N\}\big)\bigg]\bigg\} \tag{4.61}
\end{align*}

\smallskip

\noindent Where the quark-antiquark harmonic oscillator form factor coming from eqs(4.59), eq(4.54) (with for notation simplicity $\pi\alpha'=1$ and by considering the scalar case $\dfrac{dX^{\mu_1}(\sigma_1)}{d\sigma_1}\linebreak \dots \dfrac{dX^{\mu_N}(\sigma_N)}{d\sigma_N} \to 1$) is given explicitly by the result:
\begin{align*}
&\qquad\qquad F\big((P_k^\mu \dot P_{k'}^\mu), A,t,\{\sigma_1,\dots,\sigma_\mu\}\big)\,=\\
&= \left(\frac{(3+2A) \sqrt{\frac 2A}}{6\pi\,\sin h \big(\sqrt{\frac 2A}\,t\big)}\right)^{D/2}
\times \exp\bigg\{-\frac{3\sqrt A}{\sqrt 2(3+2A)\sin h\big(\sqrt{\frac 2A}\,t\big)}\\
&\times \left[\sum_{\substack{k=1\\k'=1}}^N (P_k^\mu\cdot P_{k'}^\mu)\left(\sin h \left(\sqrt{\frac 2A}\right)(t-\sigma_k)\right) \times \sin h\left(\sqrt{\frac 2A}\right) \sigma_{k'}\right]\bigg\} \tag{4.62}
\end{align*}

\smallskip

It is very important to remark that our ``toy model'' given by eq(4.62) has the correct structure to generate a Lorentz-invariant scattering amplitude, after continuation to Minkowski space, on the light of the Hall and Wightman theorem ([9]) $(2\pi\alpha'=1)$ a very important results obtained from these partially phenomenological studies on strings for QCD $(SU(\infty))$. Namely:
\begin{equation}
G_{(\Lambda_{(QCD,t)}}\, \big(P_1^\mu,\dots,P_N^\mu\big) = F_{(\Lambda_{QCD})}\,\big(P_\mu^i \cdot P_\mu^k\big) \tag{4.63}
\end{equation}

\smallskip

One point now worth to be called the reader atteention for is that case should be taken in applying straightforwardly the Feynman path integral eq(4.50) to represent the propagator of a particle possessing fermionic degrees in the presence of an external Gauge field ([3]). One can avoid this operational path integral procedure by squaring the fermionic determinant and making use now of the well-defined proper-time formalism for bosonic caloured particles ([3]). Namely (see eq(4.51))
\begin{align*}
&\qquad\qquad \det\,^{\frac 12}(\not\!D(ig A+J)\not\!D(ig A+J)\,=\\
&\qquad\qquad = \det\,^{\frac 12} \big(D^*(ig A+J)D(ig A+J)\big)\,\times\\
&\times \det \left(\pmb{1} - \frac{ig}{4}\, [\gamma^\mu,\gamma^\nu](D^*D)^{-1}(ig A+J) \times F_{\mu\nu}(ig A+J)\right) \tag{4.64}
\end{align*}

\smallskip

Since the Klein-Gordon bosonic propagator can be written in term of the $SU(N)$ normalized holonomy factor as of as
\begin{align*}
&\qquad\quad \big(D^*(ig A_\mu+J_\mu) D(ig A_\mu + J_\mu)\big) (x_1,x_2)\,=\\
&= N \left\{\int_0^\infty dt \int_{X(0)=x_1}^{X(t)=x_2} D[X(\sigma)]\,e^{-\frac 12\,\displaystyle\int_0^t \overset{\centerdot}{X}^2(\sigma)\,d\sigma}\right\}\\
&\qquad\qquad \times \psi_{x_1x_2} [C,A] \times \Phi_{x_1x_2} [C,J] \tag{4.65}
\end{align*}

\smallskip

Here the non-abelian dynamical and abelian vectorial external sources phase factors are defined explicitly by
\begin{align*}
&\psi_{x_1x_2}[C,A] = \left[\frac 1N\, \mathbb{P}\left(\exp\,ig \int_0^t A_\mu(X(\sigma))dX^\mu(\sigma)\right)\right]\\
&\Phi_{x_1x_2} [C,J] = \left[\exp\left(\int_0^t J_\mu(X(\sigma))dX^\mu(\sigma)\right)\right] \tag{4.66}
\end{align*}

\smallskip

The final expression for the generating functional eq(4.51) at large $N$, is thus easily written in the proper-time formalism, before taking the Yang-Mills path integral average is 
\begin{align*}
&\ell g\left\{\det\,^{\frac 12} \big(\not\!D^*(ig A+J)\not\!D(ig A+J)\big)\right\}\\
&= -\frac 1N \bigg\{\int_0^\infty \frac{dt}{t} \int d^4\,x_1 \int d\mu[C_{x_1x_2}]\\
&\qquad\qquad \left({\rm Tr}_{SU(N)}\,\Psi_{x_1x_2} [C,A]\right) \Phi_{x_1x_2} [C,J]\bigg\}\\
&- \bigg\{\sum_{n=2}^\infty \left(\frac 12\right)^{n-1}  \left(\frac{1}{1/N}\right)^n\\
&\int d^4x_1\dots d^4x_n \int_0^\infty (dt_1\dots dt_n) \int d\mu[C_{x_1x_2}]\\
&d\mu[C_{x_nx_1}] \times 
{\rm Tr}_{\rm Dirac}\big([\gamma_{\mu_1},\gamma_{\nu_1}]\dots[\gamma_{\mu_n},\gamma_{\nu_n}]\big)\\
&\times {\rm Tr}_{SU(N)} \bigg\{\frac{\delta}{\delta\sigma_{\mu_1\nu_1}(x_1)}\,\left(\psi_{x_1x_2} [C,A] \Phi_{x_1x_2} [C,A]\right)\\
&\qquad\qquad \cdots\, \frac{\delta}{\delta\sigma_{\mu_n\nu_n}(x_n)}\,\left(\psi_{x_nx_1} [C,A] \Phi_{x_nx_1} [C,A]\right)\bigg\}\bigg\} \tag{4.67}
\end{align*}

\smallskip

Here the Migdal-Makeenko loop derivative is introduced ([3])
\begin{equation}
\frac{\delta}{\delta\sigma_{\mu\nu}(X(\overline\sigma))} = \lim_{\varepsilon\to0} \int_{-\varepsilon}^{\varepsilon} d\zeta.\zeta \frac{\delta^2}{\delta X_\mu(\overline\sigma+\frac{\zeta}{2})\delta X_\nu(\overline\sigma-\frac{\zeta}{2})} \tag{4.68}
\end{equation}

\smallskip

\noindent which by its turn has the geometrical meaning of dividing the path trajectory $C_{x_1x_1}$, quite closely analogous to the joining and splitting picture of the old theory of dual strings (after taking the $SU(\infty)$ limit into our constant gauge fields model as given by eq(3.31) or eq(3.35) in section 3 of this paper).

\medskip

Another point worth to call attention is the expansion parameter on eq(4.67) is the color $N$, but appearing now as a Laurent power series on $\left(\dfrac{1}{1/N}\right)^{+n}$ (note that $\dfrac 1N \sim 0$, for large $N$).

\bigskip

\noindent{\bf Conclusion:} We can see from this work that another time in $QCD$ physics it is raised hopes that on underlying string dynamics is in the way to handle correctly the mathematical -- calculational aspects of Euclidean -- abelian Gauge theories in theirs confining phase, signaled here by the explicit hypothesis of a non-vanishing energy for the non-perturbative vacuum $\big(\langle 0|tr(F^2)|0\rangle \ne 0\big)$.

\medskip

At this point let us remain that our string representation for the $QCD$-Eguchi-Kawai reduced model is a free bosonic one. However if one consider next non-constant full space-time variable corrections/fluctuations to the gauge connections entering into the full Yang-Mills path integrals, one is lead to the self-avoiding fermionic full structure of the $QCD(SU(\infty))$ ([3]) with the extrinsic string as an effective bosonic string representation for $QCD(SU(\infty))$.

\medskip

Finally, we should roughly say that our path integral is at $SU(\infty)$, but surely we are in the context of a somewhat $\dfrac 1D$ expansion for the pure quantum Yang-Mills field, with a non perturbative vacuum. Unfortunately, the famous $\dfrac 1D$ expansion of Lattice $QCD$ has not been generalized or even well-understood on the continuum. We hope that our work should be a step in this direction.

\medskip

\noindent\textbf{Acknowledgments:} This research was completed under the financial support of a CNPq Visiting Senior Fellowship. The author is also thankfull to Professor W. Rodrigues -- IMECC/UNICAMP for discussions and support.

\newpage

\centerline{\large\bf Appendix A}

\vglue .3in

Let us consider the term
\begin{equation}
J_1 = \exp\left\{\frac 14\,G_0^a\,G_1^b\,G_0^c\,G_1^d\,\mathcal{L}_{abcd} \times \left[g^2V + \frac{(g^2S)^2}{(N/2)}\right]\right\} \tag{A.1}
\end{equation}

\smallskip

\noindent After the re-scaling
\begin{equation}
G_{0,1}^f = \widetilde{G}_{0,1}^f\,\left[g^2V + \frac{(g^2S)^2}{(N/2)}\right]^{-\frac 14} \tag{A.2}
\end{equation}

\smallskip

\noindent It terms out to be 
\begin{equation}
I_1 = \exp\left\{\frac 14\,\widetilde{G}_0^a\,\widetilde{G}_1^b\,\widetilde{G}_0^c\,\widetilde{G}_1^d\,\mathcal{L}_{abcd}\right\} \tag{A.3} \end{equation}

\smallskip

\noindent However a ``mixed'' term of the form
\begin{equation}
I_2 = \exp\left\{\frac 14\,G_0^a\,G_2^b\,G_0^c\,G_2^d\,\mathcal{L}_{abcd} (g^2V)\right\} \tag{A.4}
\end{equation}

\smallskip

\noindent under the re-scaling
\begin{equation}
G_2^f = \widetilde{G}_2^f\,\big[g^2V\big]^{-\frac 14} \tag{A.5}
\end{equation}

\smallskip

\noindent becomes now 
\begin{equation}
I_2 = \exp\left\{\frac 14\,\widetilde{G}_0^a\,\widetilde{G}_2^b\,\widetilde{G}_0^c\,\widetilde{G}_2^d\,\mathcal{L}_{abcd}
\,\frac{(g^2V)\times(g^2V)^{-1/2}}{\big[g^2V + \frac{(g^2s)^2}{N/2}\big]}\right\} \tag{A.6}
\end{equation}

\smallskip

Note that $N \to \infty$, we have the leading asymptotic limit
\begin{equation}
\frac{(g^2V)^{1/2}}{\big[g^2V + \frac{(g^2S)^2}{N/2}\big]^{1/2}} \,\sim\, \frac{(g^2V)^{1/2}}{(g^2V)^{1/2}} \to 1. \tag{A.7}
\end{equation}

\smallskip

\noindent It is worth recall that
\begin{equation}
\prod_{a=1}^{N^2-N} \big(dG_0^a\,d G_1^a\big) = \left\{\left[g^2V+\frac{(g^2S)^2}{N/2}\right]^{-\frac{(N^2-N)}{2}}\right\}
\left(\prod_{a=1}^{N^2-N} d\widetilde{G}_0^a\,d\widetilde{G}_1^a\right) \tag{A.8}
\end{equation}
\begin{equation}
\prod_{\substack{a=1\\ \mu\ne0,1}}^{N^2-A} dG_\mu^a = \left\{(g^2V)^{-\frac{(N^2-N)}{4}(D-2)}\right\} \times 
\left(\prod_{\substack{a=1\\ \mu\ne0,1}}^{N^2-N} d\widetilde{G}_\mu^a\right) \tag{A.9}
\end{equation}

\smallskip

We added also the remark about the constant gauge field non-abelian Stokes theorem
\begin{align*}
W[C] &= {\rm Tr}_{SU(N)} \left(\mathbb P\left(e^{ig \oint_c A_\mu\,dX_\mu}\right)\right\}\\
&= \frac 1N\, {\rm Tr}_{SU(N)} \left\{\mathbb P\left(e^{ig\,F_{\mu\nu}\big(\int ds^{\mu\nu}\big)}\right)\right\}\\
&= \frac 1N\, {\rm Tr}_{SU(N)} \left\{\mathbb P\left(e^{(ig)ig[A_\mu,A_\nu]_-\,S}\right\}\right\}(\delta_{\mu 0}\,\delta_{\nu 1})\\
&\overset{N\to\infty}{\sim}\, \frac 1N\, {\rm Tr}_{SU(N)} \left\{\pmb{1} - g^2S[A_0,A_1]_- + \frac{(g^2S)^2}{2}\,[A_0,A_1]_-^2 +\dots \right\}\\
&\overset{N\to\infty}{\sim}\, \exp\left\{ + \frac{(g^2S)^2}{2N}\, {\rm Tr}_{SU(\infty)} \big([A_0,A_1]_-^2\big)\right\} \tag{A.10}
\end{align*}

\medskip

As a  last point of our Wilson loop  evaluations at large $N$ at the context of constant gauge field configurations, we point out that at $D=2$ (the two dimensional case, it is not need to consider the phenomena of the dimensional transmutation coupling constant and the evaluation above displayed leads directly to the area behaviour for the Wilson Loop. (Remark originally due Bollini-Giambiagi) ([3]). However it is important to keep in mind that such result can be obtained quite straightforwardly by using the axial gauge $A_0^a=0$, and mostly important, it shows the non-dynamical behaviour of the pure Yang-Mills quantum (perturbative) theory at two-dimensions. At $D=3$, our $SU(\infty)$-constant gauge field model yields charge  screening instead of color charge confinement (a lenght behaviour for the Wilson Loop). Finally for $D > 4$, we have a infinite volume vanishing Wilson Loop, which by its term signals that the Yang-Mills theory in $R^D$, $D > 4$ is a trivial $QFT$, in place of the usual wrong, but always argued for  non-renormalizability of Yang-Mills theory for $D>4$.

\vglue .3in

\centerline{\large\bf References}

\vglue .1in

\begin{itemize}
\item[{[1]}] Luiz C.L. Botelho - ``Critical String Wave Equations and the $QCD(U(N_c))$ String. Int. J. Theor. Phys. (2009) 48 - 2715-2725.
\item[{-}] Jun Ishida and Akio Hosoya - Progress of Theoretical Physics, vol. 62, no. 2, 1979.
\item[{[2]}] Luiz C.L. Botelho - ``The Electric Charge Confining in Polyakov's Compact $QED$ in $R^4$''. Int. J. Theor. Phys. (2009) 48 - 1695-1700.
\item[{[3]}] Luiz C.L. Botelho - ``Methods of Bosonic and Fermionic Path Integral Representations - Continuous Random Geometry in Quantum Field Theory'', Nova Science, New York - (2008).
\item[{-}] C.G. Bollini and J.J. Giambiagi - Z. PhysC - Particles and Fields 22, 257-259 (1984).
\item[{[4]}] T. Eguchi and H. Kawai - ``Reduction of Dynamical Degrees of Freedom in the Large $N$ Gauge Theory'',  Phys. Rev. Lett., vol. 48, 1063 (1982).
\item[{[5]}] Luiz C.L. Botelho - ``The confining behavior and asymptotic freedom for $QCD(SU(\infty))$ - a constant - Gauge - field - path integral analysis'', Eur. Phys. J. 44, 267-276 (2005).
\item[{[6]}] Luiz C.L. Botelho - ``Triviality - Quantum Decoherence of Fermionic Chronodynamics $SU(N_c)$ in the Presence of an External Strong $U(\infty)$ Flavored Constant Noise Field'', Int. J. Theor. Phys. (2010) 49 - 1684-1692.
\item[{[7]}] M.N. Chernodubat all - ``On chromoeletric (super) conductivity of the Yang-Mills vacuum'', avXiv: 1212.3168 v 1 [hep-ph], 13 Dec. 2012.
\item[{[8]}] Claude Itzykson \& Jean-Michel Drouffe - ``Statistical field theory'', vol. 2, Cambridge Monographs on Mathematical Phisics, 1991.
\item[{-}] Luiz C.L. Botelho - Research Article - ``Basics Polyakov's Quantum Surface Theory on the Formalism of Functional Integrals and Applications'', International Scholarly Research Network ISRN High Energy Physics, vol. 2012, Article ID674985, doi: 10.5402/2012/674985.
\item[{-}] C. Itzykson and J.B. Zuber - Nuclear Physics B275 [FSS7] 580-616. (1986).
\item[{[9]}] N.N. Bogoluhov, A.A. Logunov and I.T. Todorov - Introduction to Axiomatic Quantum Field Theory - Mathematical Physics Monograph series - Benjamin/Cummings Publishing Company Inc. - USA, 1978.
\end{itemize}

\end{document}